# Revealing essential dynamics from
# high-dimensional fluid flow data and operators

＊カリフォルニア大学ロサンゼルス校　工学部　機械航空宇宙工学科　　　平　　　邦　　彦 †

Mechanical and Aerospace Engineering, University of California, Los Angeles　　Kunihiko Taira

## 1 Introduction

We have seen tremendous developments in computational fluid dynamics (CFD) solvers over the past few decades. These CFD solvers have empowered engineers and scientists to analyze a variety of fluid flows and established themselves as a critical component of engineering design processes for fluid flow systems[3]. Together with the advancement in solver algorithms, the continuous enhancement in computational resources is easing the challenges of analyzing complex fluid flows with high dimensionality and multi-scale nonlinear dynamics[1,2].

With the increase in computational speed and availability of large disk space, numerical simulations have been generating enormous amount of data. It has become a tremendous task to go through the data and extract physical insights contained therein based on traditional approaches of visualizations and statistical assessments. In fact, for some massive computations, post processing and analysis have become a mission in itself requiring a team of scientists and engineers. Even with the best intentions, we are now at the point where some parts of large CFD data may be left unanalyzed. Without doubt, there must be countless high-fidelity CFD data on precious hard drive space that have been untouched or simply forgotten. As we further advance our computational capabilities, we will likely enter an era where post processing and detailed analysis of fluid flow data cannot be easily performed without appropriate mathematical tools.

These trends are not unique to CFD. Experimental diagnostic tools collect intricate measurements of the flow in two and three-dimensional manners, yielding large data sets[4]. Moreover, the management and analysis of large-data sets are critical issues in many fields, including natural science, computer science, and data science. While the data structures are generally unique to each research field, there are a number of commonalities among different fields to allow for cross-pollination of foundational concepts in analyzing data containing dynamical information. The time to look into these issues is ripe with tremendous developments in data-inspired techniques emerging from data science and applied mathematics.

In addition to the flow field data that we collect from simulations, a wealth of knowledge is contained in the governing Navier-Stokes equations. We can consider analyzing the Navier-Stokes operator linearized about some base flow. The linearized Navier-Stokes operator for a chosen base flow can be constructed with appropriate discretizations and boundary conditions[5,6]. Depending on the Reynolds number of the flow, the discrete operator can become very large. Although the analysis of the operator can be computationally demanding, the insights extracted from such operator can provide details on the flow field, such as stability and input-output properties[7]. Knowledge gained from operator-based analysis can also benefit from the latest development in data science, as handling of large operators are becoming a critical need for the whole research community, beyond just fluid mechanics.

To address the aforementioned challenges, we consider, in this paper, concepts centered around modal analysis, data science, network science, and machine learning to reveal the essential dynamics from high-dimensional fluid flow data and operators. The presentation of the material herein is example-based and follows the author's keynote talk at the *32nd Computational Fluid Dynamics Symposium* (Japan Society of Fluid Mechanics, Tokyo, December 11-13, 2018). This talk was delivered as a compilation of some of the research activities undertaken by the author's research group. Due to space limitations, we only present a

*48-121 Engineering IV, UCLA, Los Angeles, CA 90095-1597

†E-mail: ktaira@seas.ucla.edu

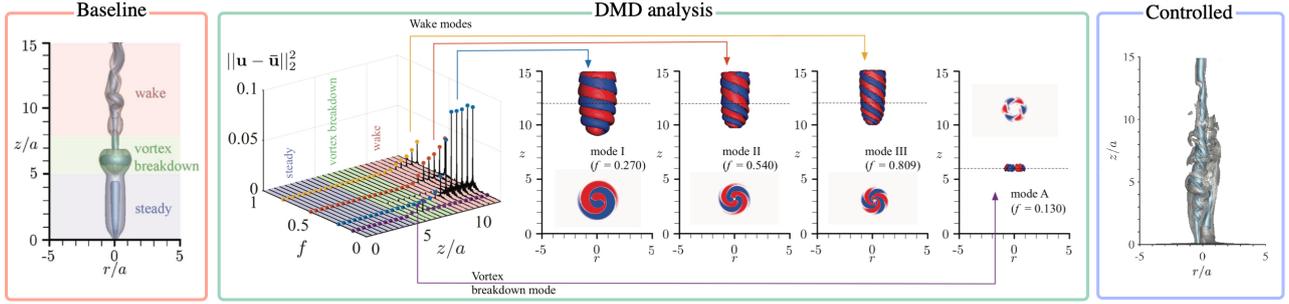

Fig. 1 DMD-based analysis of an unsteady wall-normal vortex at $Re = 5\,000$ and the application of active flow control to increase the vortex core pressure[21]. Reprinted with permission from Cambridge University Press.

brief overview of these topics and kindly direct the readers to the cited references. We hope that the discussions shown below stimulate the curiosities of the readers to peek into some of these new fields if they have not already.

## 2 Modal analysis

Modal analysis can extract physically important spatial modes from fluid flows[8]. The analysis can be broadly categorized into data-based and operator-based techniques. The data-based techniques include the celebrated proper orthogonal decomposition (POD)[9–13] and dynamic mode decomposition (DMD)[14–16], which can determine energetically and dynamically important modes from a collection of instantaneous flow fields. The operator-based techniques, including the global stability analysis[6,17] and the resolvent analysis[7,18], require a base flow and the discrete linearized Navier-Stokes operator. For high-Reynolds number flows with complex dynamics, the fluid flow data set and the Navier-Stokes operator become very large due to the dimensionality of the dynamical system. In this section, we present the use of dynamic mode decomposition (DMD) and resolvent analysis on two examples of unsteady flows at moderate Reynolds numbers for their characterizations and control.

### 2.1 Data-based analysis (DMD)

The emergence of strong vortices upstream of a liquid pump can be detrimental to its performance and safety[19]. This is especially critical when the vortex becomes so strong with an unsteady hollow core, caused by the low-pressure core accumulating air bubbles in the free stream or from the free surface[20]. As the first example, we consider modifying the characteristics of a wall-normal vortex based on the insights from DMD analysis[21].

We model this type of vortex by prescribing the Burg-
ers vortex inflow velocity profile but specify a no-slip wall at the bottom of the domain. The swirl flow enters from the side and exits the domain from the top, as shown in Figure 1 (left). The flow generates a strong unsteady vortex at a circulation based Reynolds number of $5\,000$. In order to understand the unsteady behavior of this vortical flow, we use DMD to reveal the spatial structures of the dominant unsteadiness. We collect the snapshots of the velocity field from direct numerical simulation and extract the spatial modes using DMD. The dominant DMD modes are visualized in Figure 1 (middle) with the corresponding frequency components, which can be determined from the DMD eigenvalues.

By examining these modes and frequencies, we can elucidate that unsteadiness held by certain modes are beneficial in increasing the vortex core pressure. With the gained insights, we introduce unsteady actuation (blowing and suction) from the bottom wall at the vortex center with an appropriate frequency and azimuthal wavenumber[21]. The chosen control input is able to trigger the emergence of instabilities farther upstream, which leads to the widening of the vortex core and reducing the azimuthal velocity of the vortex, as visualized in Figure 1 (right). This in turn causes the vortex core pressure to be increased, which could help avoid detrimental effects if the vortex is to be engulfed by a hydrodynamic pump downstream.

### 2.2 Operator-based analysis (resolvent analysis)

There are two major operator-based modal analysis techniques for fluid flows. First is the global stability analysis, which is an eigenvalue analysis that determines the stability characteristics of the flow about an equilibrium state. This analysis essentially serves as a perturbation analysis in terms of an initial value problem. In contrast, there is the resolvent analysis, which examines the harmonic response of a system about a given base state. This latter approach examines the harmonic input-output relationship for a flow and can use either an equilibrium or a time-averaged base flow[18]. The resolvent analysis reveals the characteristics of the particular solution for a continuously forced system. Here, we present the use of the resolvent analysis to design the flow control strategy for suppressing turbulent flow separation over an airfoil[22].

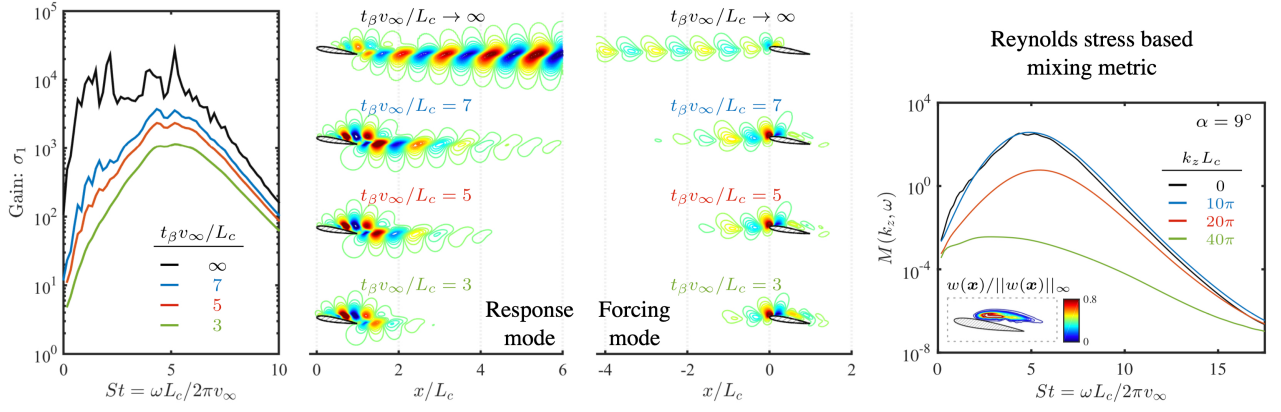

Fig. 2　Resolvent analysis of turbulent separated flow over an NACA 0012 airfoil[22]. Shown are the gain distribution, response and forcing modes (vertical velocity component visualized at $St = 0.833$), and the modal mixing metric based on the Reynolds stress. The finite-time window is denoted by $t_\beta$, normalized by the freestream velocity $v_\infty$ and chord $L_c$. Reprinted with permission from Cambridge University Press.

In this example, we consider a turbulent separated flow over a NACA0012 airfoil at $Re = 23\,000$ and $\alpha = 9°$. The flow is taken to be spanwise periodic and simulated with large eddy simulation. The time averaged base flow is used to construct the resolvent operator that relates the forcing input to the system output for a chosen frequency. When the flow is in a statistically stationary state, the nonlinear terms in the Navier-Stokes equations can be viewed as the forcing input, which makes the resolvent analysis a valuable tool to study turbulent flows[18]. Since this turbulent base flow is unstable, we note that the input-output analysis must be conducted on a time scale shorter than those of the instabilities[22,23]. This perspective enables us to determine possible amplification path that influences the base state even in the presence of the inherent instabilities.

The resolvent analysis is performed by taking the singular value decomposition of the *resolvent operator*, which maps the forcing input to the response output. The left and right singular vectors of the resolvent operator respectively correspond to the response and forcing modes, and the singular values represent the gain. That is, for a temporally oscillating input at a certain frequency with the forcing mode profile, that input is amplified (or attenuated) by the gain in the form of the response mode. The dominant forcing and response modes are shown in Figure 2 along with the gain distribution over the frequency. Also shown is the influence of the finite-time window (discounting).

To suppress separation over the wing, we aim to enhance turbulent mixing by introducing a number of unsteady flow control actuators in the spanwise direction near the separation point. In this particular study, we consider the use of a thermoacoustic actuator[24–26]. Instead of performing a large number of parametric LES computations, we can use the knowledge from resolvent analysis to find an effective flow control setup. We do so because the resolvent analysis is considerably inexpensive compared to the full LES calculation. We can take the dominant resolvent modes and evaluate the dominant Reynolds stress to assess the level of turbulent mixing to energize the separated region of the flow. Shown on the right is the mixing metric based on the spatial integral of the spanwise Reynolds stress that identifies the optimal combination of actuation frequency and spanwise wavenumber. These control parameter choices are verified with LES for its effectiveness in suppressing separation for this turbulent flow example. The use of resolvent analysis together with physical insights to derive a metric for developing an active flow control strategy can thus serve as a powerful tool.

## 3　Network science

Over the past decade, network science has become a very active field of research that studies the structure and dynamics of interactions[27–32]. Network based analysis can examine interactions that take place over a collection of elements, which is mathematically described by a graph. The graph structure of interactions is examined with graph theory[33], and networked dynamics on a graph can be analyzed by incorporating dynamical systems and control theories[34,35]. The strength of network science lies in its ability to examine a range of networked problems in epidemiology, biology, engineering, computer science, cyber security, and sociology, amongst many others, enabling refreshingly new concepts to come together. In recent years, network analysis has been extended to fluid dynamics to examine vortical interactions[36,37], turbulent interactions[38,39], modal interactions[40], and cluster networks[43]. We briefly describe some of our data-inspired efforts, in particular the modal and cluster-based network analyses.

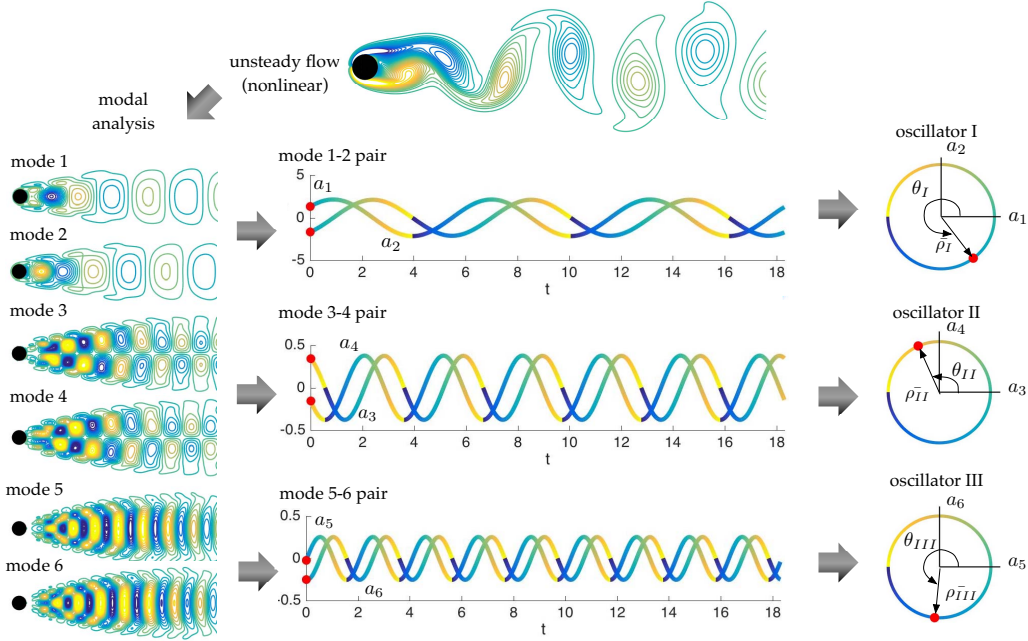

Fig. 3   Modal oscillator representation of two-dimensional cylinder flow at $Re = 100$, for which POD mode pairs constitute oscillators. Reprinted with permission from the American Physical Society.

### 3.1   Modal network

To start the network-based analysis of fluid flows, let us consider the two-dimensional laminar cylinder wake at $Re = 100$. In particular, we examine the kinetic energy transfer among the modal structures[40]. We can first find the POD modes from a collection of instantaneous snapshots of the velocity field using the snapshot POD method[11]. Since the wake is convective and periodic in nature, the modes appear in pairs. In fact, the coefficient of each mode pair can be compactly expressed as an *oscillator* using a complex variable representation, as illustrated in Figure 3.

In the baseline oscillatory state, the oscillators (mode pairs) maintain a constant level of energy. However, when the flow is perturbed, these modal oscillators transfer perturbation kinetic energy among themselves over the modal interaction network. We can describe such interactions using the networked Stuart-Landau model and determine the transfer coefficient (weighted network) through a data-based least squares process[40]. The revealed *modal oscillator network* is shown in the blue box of Figure 4. The perturbation energy is distributed over the network based on what is referred to as the graph Laplacian. The shown network visualizes the flow of kinetic energy. It can be seen that the first oscillator (mode pair) is dominating in passing down the energy without getting influenced from the others. This oscillator can be regarded as the leader on the network and the rest of them being followers[34].

This modal-network model is a linear model and can be seamlessly used with modern control toolsets. As stabilization of the wake can lead to drag reduction, we consider the use of body force actuation and develop a linear quadratic regulator based feedback control to attenuate the fluctuations in the flow. The controlled dynamics lead to the modified network shown in the green box of Figure 4, in which we have modified the energy transfer path. This in turn leads to the reduction in unsteady fluctuations as well as drag. This result is summarized for controllers with different gains on the manifold comprised of the dominant mode coefficients and the shift mode[41], which represents how far the flow is from the unsteady equilibrium point (minimum drag). This type of network model not only provides an intuitive guide to understand flow physics but also accurately models the energy transfer path, which has been a difficulty with the traditional Galerkin projection models[40].

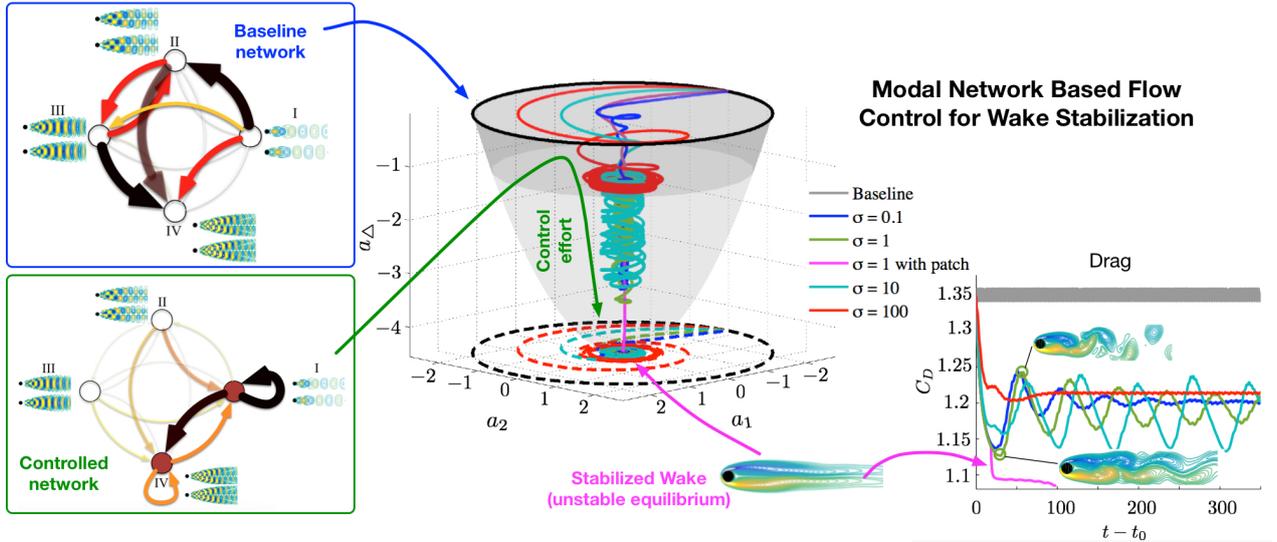

Fig. 4   Modal-interaction network for a two-dimensional cylinder flow. Feedback control is performed on the networked dynamics and is shown to yield wake stabilization and drag reduction[40]. Reprinted with permission from the American Physical Society.

### 3.2   Cluster network

We can extend the concept of networks and take another data-based approach to characterize the dynamics of the flow. Here, we take inspiration from the work of Kaiser et al.[42] on cluster-based control and develop a cluster-based network formulation. The innovative approach by their work generates a probabilistic model using the so-called *feature space* to discretize the dynamics of the flow. For example, one can consider studying the trajectory of the flow dynamics in a space spanned by the dominant POD temporal mode coefficients. The trajectory can then be discretized into *clusters* that partitions the features space. The presence of the flow state and its transition from one cluster to another can be modeled in a probabilistic manner.

Here, let us present the cluster based discretization[43] of the turbulent flow over a NACA 0012 airfoil at $Re = 23\,000$. In this example, we take the feature space to be defined by $C_L$, $\dot{C}_L$, and $C_D$. These force coefficients are chosen to characterize the lift and drag variations as well as the phase dynamics (captured by $C_L$ and $\dot{C}_L$). Clustering can be performed with the K-means algorithm as shown in Figure 5 (left). The volume occupied by each cluster can be reduced to its centroid for modeling and the state transition from one cluster to another can be described by a probabilistic transition matrix, visualized in Figure 5 (top right). Community detection techniques can be applied to study this feature space, from which we find three main communities in this example. They correspond to low, intermediate, and high drag states, as illustrated in Figure 5 (bottom right).

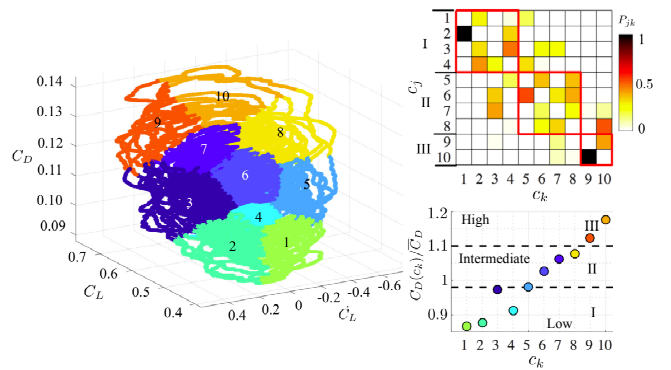

Fig. 5   Clustering of feature space for unsteady 3D separated turbulent flow over a NACA 0012 airfoil at $Re = 23\,000$.[43] Reprinted with permission from Cambridge University Press.

With the cluster network established, we pose an optimization problem of determining the best active control setup (with time-varying blowing near the leading edge) that minimizes the aerodynamic power consumption, including the actuation cost. That is, we seek the optimal control amplitude at each cluster state, but in a globally coupled manner for all clusters. To solve this problem numerically, an iterative optimizer based on a simplex method is applied. This whole optimization routine involves LES calculations in each of the iterative steps but shows convergence to the optimal solution in $\mathcal{O}(10)$ iterations.

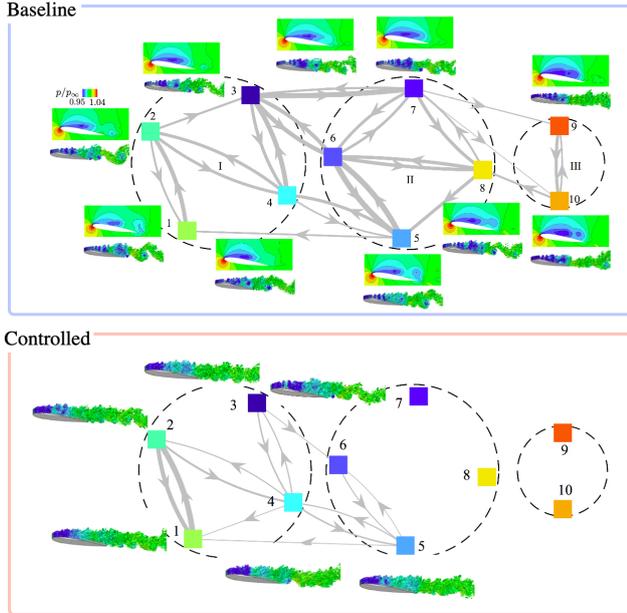

Fig. 6   Cluster network (Markov chain) visualized for the baseline and optimally controlled flows over a NACA 0012 airfoil at $Re = 23\,000$.[43]. Active flow control is applied to modify the cluster network for minimal aerodynamic power consumption. The three circles indicate the low, intermediate, and high-drag communities (from left to right). Reprinted with permission from Cambridge University Press.

Compared in Figure 6 are the baseline and optimally controlled probabilistic cluster network (Markov chain). The controlled flow achieves $13\%$ drag reduction. Noteworthy here is that the algorithm does not necessarily find a control setup that eliminates separation, as we have not specified that as part of the cost function. Hence, it is possible that this type of optimizers can find control techniques that are not traditionally examined. This type of control leverages data-based characterization of high-dimensional flow without having to perform a large number of parametric investigations[44], which is computationally expensive. While the present approach is still computationally heavy, smart extraction of insights from large-data may further reduce the iteration time to find optimal solutions for various fluid mechanics problems.

## 4   Emerging methods

In this last section, let us discuss two emerging techniques from data science that I think can help advance the analysis techniques for fluid mechanics. I will in particular bring up *machine learning* and *randomized methods*, which we have tried in our research group. Based on our efforts with these techniques, we find them to hold great potentials within the overall goal of this paper and beyond. We also will discuss some of the open questions associated with these techniques.

### 4.1   Machine learning

In recent years, machine learning has become a very popular tool to tackle a variety of fluid flow problems, especially with increased computational resources (especially, the GPUs) and access to the machine learning libraries. In the field of fluid mechanics, machine learning techniques have demonstrated their strengths in a variety of applications, including turbulence modeling[45–49], reduced-order modeling[50], classification[51,52], and flow control[53,54]. Successful uses of machine learning are reported in a large number of studies and we expect the number of such reports to grow at a rapid pace.

Machine learning algorithms for developing models require access to a large collection of reliable reference data for the machine to learn the trend in the data. For example, machine learned models can take a form of $y = F(x; w)$, where $x$ is the input and $y$ is the output for model $F$ with parameters $w$. In this case, we can consider the determination of the weights $w$ to be a regression problem and $F$ can be constructed over various structures (e.g., neural networks). These machine based models can be a great regression tool and can be applied to a variety of fluid flow problems. It should be noted that all machine learning algorithms cannot be simply expressed in one way as it has a broad definition.

In the case of a regression type problem, one of the open questions is how one should choose a structure on which a model is constructed. There are many choices that one can make from a large family of neural networks[55]. Even after one selects a network, the size of the network needs to be appropriately selected. At the moment, there appears to be a lack of constructive guidance on how the selection should be made for fluid dynamics problems. The other important thing to note here is that machine learned models are great for interpolation against the training data. However, their solutions can be grossly off if the models are asked to extrapolate. In other words, predictions are difficult – a machine learned model cannot accurately reproduce a solution outside the span of the training data. What makes this further difficult is that the distinction of interpolation and extrapolation can be vague for machine learning. For this reason, prediction of dynamics remains to be a challenge, unless certain features of the dynamics lie in some state space that the machine has seen before.

Let us present here a successful regression type machine learning example. Here, we demonstrate the use of a machine learned super-resolution analysis to reconstruct two-dimensional isotopic turbulence from a grossly under-sampled image. The training data is provided from high-resolution DNS and the input data is provided by under sampling the flow field. By training the model over a hybrid neural network, we find that even with a very low resolution image (with $4 \times 4$ pixels), the turbulent flow field can be reconstructed with striking fidelity, as shown in Figure 7. This shows that machine learning models can perform complex interpolations based on the given training data to reproduce the global flow field. This opens possibilities to reproduce a variety of complex fluid flows, including three-dimensional turbulence with sufficient training. While not shown, the same approach performs very well for super-resolution reconstruction of two-dimensional laminar flows[57].

As machine learning based techniques are now being tried for a wide variety of problems in fluid mechanics, we should start to identify successes and challenges in incorporating the concepts into our analytical toolsets. One of the main challenges that is already identified with machine learning is the interpretability of results. Other questions include how one should collect the training data, as fluid mechanics data sets are different in their properties from what data scientists generally handle. As we make progress, machine learning based techniques should serve as invaluable tools in fluid mechanics.

## 4.2 Randomized methods

As we pursue the analysis of three-dimensional higher Reynolds number flows with complexity, the size of the operator that describes flow physics also increases due to the necessary grid resolution. This leads to high computational cost and memory requirement for operator based analysis. For this reason, resolvent analysis for practical flow applications becomes possible only for those with significant computational resources. The computational cost has made the use of resolvent analysis (as well we other modal analysis techniques in general) limited to low Reynolds number flows in many cases.

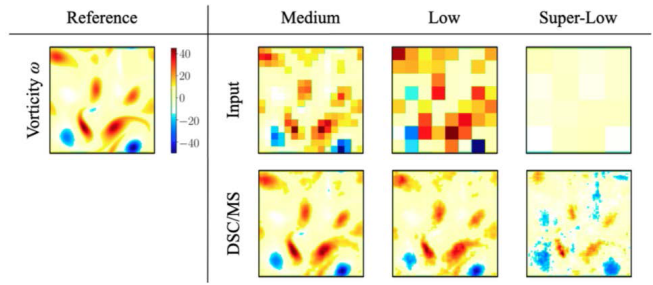

Fig. 7 Super-resolution analysis of two-dimensional turbulent flow[56]. (Top) Coarse input fields with varied resolution. (Bottom) Super-resolution reconstruction with the Down-sampled Skip-Connection/Multi-Scale (DSC/MS) model. Reprinted with permission from Cambridge University Press.

In recent years, there have been exciting developments in *randomized numerical linear algebra*[58,59]. Randomized methods have been applied to POD[60], DMD[61], and resolvent analysis[62,63], achieving speed up and memory relief. Randomized techniques are based on the idea of deriving a low-rank approximation of a large matrix, with an error bound that holds most of the time. This means that the numerical technique can achieve a significant speed up with accuracy that holds with high probability.

Here, we present how resolvent analysis can benefit from the randomized techniques, using the above example from Section 2.2. To implement the randomized technique, a randomly generated tall and skinny test matrix can be passed through the resolvent operator to extract the dominant effects and to find its low-rank representation. While SVD for the full resolvent operator can be very computationally expensive and memory intensive, a smaller size SVD can be performed on the reduced matrix to find the dominant forcing and response modes as well as the gain[62].

For example, we can reduce the resolvent operator of size 0.75 million by 0.75 million (from Section 2.2) to 0.75 million by 10. With the randomized method, computation can be accelerated without sacrificing the accuracy of the results, as shown in Figure 8. This type of analysis is especially beneficial as we do not need to find all modes but only the physically important ones. In many cases, we only seek one or two of these resolvent modes, which makes randomized methods an ideal approach. Innovations such as the randomized method has the potential to change the landscape of various analysis techniques for fluid mechanics.

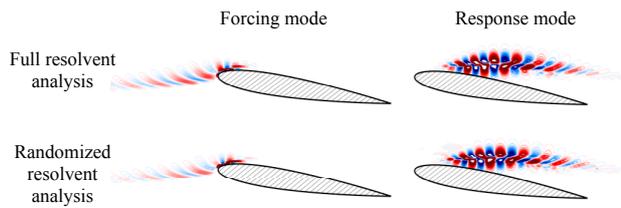

Fig. 8    Randomized resolvent analysis revealing the dominant forcing and resolvent modes accurately.[62] The full analysis is based on an operator of size 0.75 million × 0.75 million, while the resolvent analysis is based on an operator of size 0.75 million × 10. Reprinted with permission from Cambridge University Press.

## 5   Concluding remarks

The material contained in this paper is a summary of the author's keynote presentation at the 32nd Computational Fluid Dynamics Symposium in Tokyo. As many data science and applied mathematics techniques empower us to analyze vast data sets and large-scale problems, we fluid mechanicians enter an exciting era to make headways into uncovering insights on flow physics, deriving new models, and developing technologies to control the behavior of high-dimensional fluid flows with complex dynamics.

The materials presented here were based on the work of the author's research group. While they are part of our ongoing research activities, we hope they provide some stimulating developments that can complement traditional analysis techniques to expand the horizon of fluid flow analysis. This paper was compiled as an assortment of research tapas and was not intended to be a paper with detailed descriptions. If any of these topics tickled the readers' appetite, we invite the readers to take a peek into some of the references for their ingredients and recipes.

## Acknowledgements

The author acknowledges the generous support from the US Air Force Office of Scientific Research, US Army Research Office, US Office of Naval Research, and Ebara Corporation. He also thanks the insightful discussions with Steve Brunton, Nathan Kutz, Kozo Fujii, Koji Fukagata, Bernd Noack, Peter Schmid, Omer San, and Toshiyuki Arima. This paper was made possible with the contributions over the years from many students, postdocs, and collaborators, including Chi-An Yeh, Aditya Nair, Qiong Liu, Kai Fukami, Jean Helder Marques Ribeiro, Eurika Kaiser, Byungjin An, Motohiko Nohmi, and Masashi Obuchi. At last, the author thanks the organizers of this symposium for extending their invitation to deliver this keynote talk.